\documentclass[twocolumn,showpacs,pra]{revtex4}
\usepackage{graphicx}
\usepackage{dcolumn}
\usepackage{bm}
\begin{document}

\title{Marcus Electron Transfer Reactions with Bulk Metallic Catalysis}
\author{A. Widom and J. Swain}
\affiliation{Physics Department, Northeastern University, Boston MA USA}
\author{Y.N. Srivastava}
\affiliation{Physics Department, University of Perugia, Perugia IT}

\begin{abstract}

Electron transfer organic reaction rates are considered employing the
classic physical picture of Marcus wherein the heats of reaction are
deposited as the energy of low frequency mechanical oscillations of
reconfigured molecular positions. If such electron transfer chemical
reaction events occur in the neighborhood of metallic plates, then
electrodynamic interface fields must also be considered in addition to 
mechanical oscillations. Such electrodynamic interfacial electric fields 
in principle strongly effect the chemical reaction rates. The thermodynamic 
states of the metal are unchanged by the reaction which implies that 
metallic plates are purely catalytic chemical agents.

\end{abstract}


\maketitle

\section{Introduction \label{intro}}

\begin{figure}
\scalebox {0.5}{\includegraphics{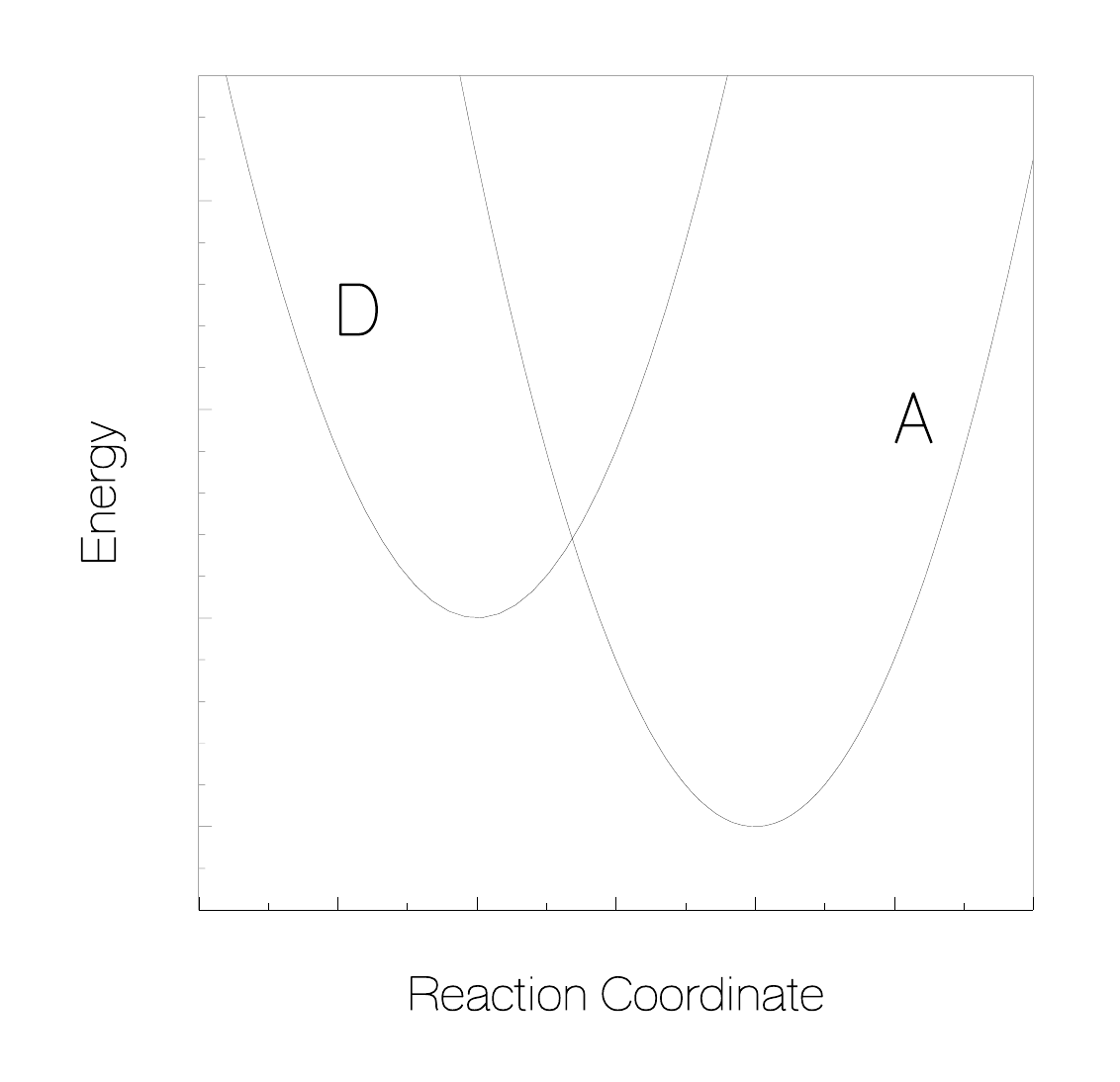}}
\caption{Shown in a schematic fashion are typical diabatic energy
curves for {\it D}-states and {\it A}-states describing, respectively,
the chemical states $DA$ and $D^+A^-$. The adiabatic energy curve is
the minimum of the two diabatic values. The energy barrier for the
reaction in Eq.(\ref{intro1}) is at that reaction coordinate wherein
the two diabatic energy curves meet.}
\label{fig1}
\end{figure}

We consider so called {\em electron transfer} chemical reactions of
the following sort. One starts from a molecule that has two parts
\begin{math} D \end{math} and \begin{math} A \end{math} known as
the donor part and the accepter part. A single electron passes from
the donor  \begin{math} D \end{math} to the accepter
\begin{math} A \end{math} according to
\begin{equation}
DA \ \longrightarrow \ D^+ A^- .
\label{intro1}
\end{equation}
After the electron transfer, the atomic positions within the molecule
reorganize so as to minimize the final free energy. If one writes
\begin{math} {\cal H}_D \end{math} as the Hamiltonian for the molecular
configuration \begin{math} DA \end{math} before the electron transfer and
\begin{math} {\cal H}_A \end{math} as the Hamiltonian for the molecular
configuration \begin{math} D^+A^- \end{math} after the electron transfer,
then the total model Hamiltonian may be written in the form
\begin{eqnarray}
{\cal H}=\frac{1}{2}\left({\cal H}_D+{\cal H}_A\right)+
\frac{1}{2}\left({\cal H}_D-{\cal H}_A\right)\sigma_3
+\hbar \Omega \sigma_1\ ,
\nonumber \\
{\cal H}=\pmatrix{{\cal H}_D &  \hbar \Omega
\cr \hbar \Omega & {\cal H}_A} ,
\label{intro2}
\end{eqnarray}
wherein \begin{math} (\sigma_1,\sigma_2,\sigma_3) \end{math} are the Pauli
matrices describing the operators for the two dimensional quantum state space
for the electron.

The physical chemical kinetic picture of how this Hamiltonian works was
pioneered by Marcus\cite{Marcus:1956,Marcus:1985,Marcus:1993,MarcusN:1993} 
and further developed by many other
workers\cite{Jortner:1976,Finklea:2001,Migliore:2012,Bazant:2013,Bai:2014}.
The physical situation\cite{Krishtalik:1986} is shown schematically in 
FIG. \ref{fig1}. The stable free energy curves as a function of reaction 
coordinates may be defined as
\begin{eqnarray}
\hat{\cal G}=\hat{\cal G}_D\ \ \ {\rm if}
\ \ \ \hat{\cal G}_D < \hat{\cal G}_A ,
\nonumber \\
\hat{\cal G}=\hat{\cal G}_A\ \ \ {\rm if}
\ \ \ \hat{\cal G}_A < \hat{\cal G}_D .
\label{intro3}
\end{eqnarray}
The maximum in the stable free energy takes place at that free energy wherein
the two different free energy curves meet,
\begin{equation}
\hat{\cal G}={\cal G}_{\rm max}\ \ \ {\rm if}
\ \ \ \hat{\cal G}_D = \hat{\cal G}_A ,
\label{intro4}
\end{equation}
at the free energy barrier upward pointing cusp. The chemical kinetic rate
of Eq.(\ref{intro1}) is thereby determined by the thermal activation flow
passing over the free energy barrier. The minima in the free energy
curves
\begin{equation}
{\cal G}_{D,A}=\min \hat{\cal G}_{D,A}
\label{intro5}
\end{equation}
yield the free energy of the reaction
\begin{equation}
\Delta {\cal G} = {\cal G}_D - {\cal G}_A .
\label{intro6}
\end{equation}
On the other hand, the free energy of the barrier over which the electron
transfer must pass is given by
\begin{equation}
{\cal B}={\cal G}_{\rm max}-{\cal G}_D,
\label{intro7}
\end{equation}
{\em i.e.} the kinetic reaction rate has the general thermal activation form
\begin{equation}
\Gamma = \nu e^{-{\cal B}/k_BT}.
\label{intro8}
\end{equation}

If the diabatic energy curves are quadratic in the reaction coordinates with
the same force constants, then the resulting oscillations merely have
their centers displaced giving rise to a reorganization potential energy
\begin{math} \lambda  \end{math}. The central result of Marcus for this
case is the barrier energy
\begin{equation}
{\cal B}_{\rm Marcus}=
\left[\frac{(\lambda -\Delta {\cal G})^2}{4\lambda }\right]
\ \ \ {\rm wherein }\ \ \ \lambda > \Delta {\cal G} > 0.
\label{intro9}
\end{equation}
The assumption of pure translations of a classical quadratic potential energy
will be dropped in much of what follows.

In Sec.\ref{rr}, the chemical reaction rate will be computed from
Eq.(\ref{intro2}) on the basis of the Fermi golden rule
\begin{math}  \Gamma=(2\pi /\hbar)|V|^2g_f \end {math}. The matrix
element in frequency units is \begin{math} |V|/\hbar = \Omega  \end{math}.
The density of final states \begin{math}  g_f \end{math} determines the
effective modulation  time scale  \begin{math} \tau =2\pi \hbar g_f \end{math}.
Thus, \begin{math} \Gamma =\Omega^2 \tau \end{math} where the condition
for the validity of the Fermi golden rule is
\begin{math} \Omega \tau \ll 1  \end{math}. The rigorous expression for
\begin{math} \tau \end{math} in terms of the Hamiltonian Eq.(\ref{intro2})
will be exhibited. In Sec.\ref{mt}, the displaced mechanical oscillators
will be explored and the Marcus calculation of the model will be discussed.
In Sec.\ref{cpb}, it is proposed that catalytic agents lower the binding 
energy of the product state in \begin{math} DA \to D^+ A^- \end{math}. The 
reason that final state binding lowers the barrier factor and thereby increases 
the reaction rate is reviewed. This is the {\em central mechanism} we are 
proposing for how metallic catalytic agents increase the rate of electron 
transfer reactions.   

In Sec.\ref{pwr}, we consider the reaction 
\begin{math} H_2O \ \longrightarrow \ H^+ OH^- \end{math} 
in pure water from the viewpoint of Marcus theory.  
The metallic catalytic rate of this pure water reaction depends 
on the ordered polarized layer of water, as discussed Sec.\ref{owl}, 
that sits on and above the metallic surface. In Sec.\ref{pw} the 
physical meaning of the experimental electric dipole moment of 
a single water molecule is explored. In Sec.\ref{ebfw}, it is shown why 
the electron energy shift is sufficiently large to totally eliminate 
the energy barrier.  

In Sec.\ref{mic}, the electron diabatic energy shift near the interface 
between a metal and an insulator is explored. The thermodynamic relations 
for the metal-insulator interface are discussed in Sec.\ref{lge}. The 
renormalization of the free energy then eliminates the barrier as explained 
in Sec.\ref{meer}. Finally, the general reasons for the metallic boundaries 
for eliminating the barrier to electron transfer reactions are discussed 
in the concluding Sec.\ref{conc}.

\section{Reaction Rate \label{rr}}

To compute the transition reaction rate
\begin{equation}
\Gamma = \Gamma_{DA \to D^+A^-} \ ,
\label{rr1}
\end{equation}
between energy eigenstates
\begin{math} \left|iD \right> \to \left|fA \right>   \end{math}
one may employ the Fermi golden rule 
\begin{equation}
\Gamma = \left[\frac{2\pi}{\hbar}\right] (\hbar \Omega)^2 
\sum_{i,f} p_{iD} \left|\big<fA \big| iD\big> \right|^2
\delta \big( {\cal E}_{fA} - {\cal E}_{iD}\big),
\label{rr2}
\end{equation}
wherein there is an average over initial states and sum over 
final states. The reaction rate may thereby be written in terms 
of the thermal electron transfer time \begin{math} \tau \end{math}, 
\begin{equation}
\Gamma = \Omega ^2 \tau .
\label{rr3}
\end{equation}
In detail  
\begin{eqnarray}
\tau = 2\pi \hbar  
\sum_{i,f} p_{iD} \left|\big<fA \big| iD\big> \right|^2
\delta \big( {\cal E}_{fA} - {\cal E}_{iD}\big), 
\nonumber \\ 
\tau = \int_{-\infty}^\infty  
\sum_{i,f} p_{iD} \left|\big<fA \big| iD\big> \right|^2 
e^{i( {\cal E}_{iD} - {\cal E}_{fA})t/\hbar} dt ,
\label{rr4}
\end{eqnarray}
yielding the final electron transit time for 
the reaction \begin{math} DA \to D^+ A^-  \end{math}; 
It is
\begin{equation}
\tau = \int_{-\infty}^\infty 
\left<e^{i{\cal H}_{D}t/\hbar}
e^{-i{\cal H}_{A}t/\hbar}\right>_D dt, 
\label{rr5}
\end{equation}
wherein the average employs the thermal canonical density operator 
\begin{math} \rho_D={\cal Z}_D^{-1} e^{-{\cal H}_D/k_BT} \end{math}  
for the initial states.

If one considers a microscopic model as in Eq.(\ref{intro2}), 
then Eqs.(\ref{rr3}) and (\ref{rr5}) are reliable for predicting 
chemical kinetic electron transfer rates provided that the motional 
narrowing condition \begin{math} \Omega \tau \ll 1 \end{math} holds 
true. Define an interaction \begin{math} {\cal V} \end{math} 
according to 
\begin{eqnarray}
{\cal H}_A={\cal H}_D+{\cal V},
\nonumber \\  
{\cal V}(t)=e^{i{\cal H}_{D}t/\hbar}{\cal V}e^{-i{\cal H}_{D}t/\hbar}.
\label{rr6}
\end{eqnarray}
From Eqs.(\ref{rr5}) and (\ref{rr6}) one finds that 
\begin{equation}
\tau =\int_{-\infty}^\infty \left<
\exp\left(-\frac{i}{\hbar }\int_0^t {\cal V}(s)ds \right)
\right>_+ dt,
\label{rr7}
\end{equation}
wherein the subscript ``\begin{math} + \end{math}'' indicates time 
ordering. 

\section{Marcus Theory \label{mt}}

The displaced oscillator model may be described by
\begin{eqnarray}
{\cal H}_D={\cal V}_D^{(0)}+\frac{1}{2}
\sum_k \big(P_k^2+\omega_k^2 Q_k^2\big),
\nonumber \\
{\cal H}_A={\cal V}_A^{(0)}+\frac{1}{2}
\sum_k \big(P_k^2+\omega_k^2 (Q_k-\xi_k)^2\big),
\label{mt1}
\end{eqnarray}
so that Eq.(\ref{rr6}) reads
\begin{equation}
{\cal V}={\cal V}_A^{(0)}-{\cal V}_D^{(0)}+
\frac{1}{2}\sum_k \omega_k^2 \xi_k^2-
\sum_k \omega_k^2 \xi_k Q_k\ .
\label{mt2}
\end{equation}
The displacement reorganization energy is thereby
\begin{equation}
\lambda = \frac{1}{2}\sum_k \omega_k^2 \xi_k^2
\label{mt3}
\end{equation}
and the interaction potential Eq.(\ref{mt2}) reads
\begin{equation}
{\cal V}=(\lambda -\Delta {\cal G})
- \sum_k \omega_k^2 \xi_k Q_k\ .
\label{mt4}
\end{equation}
The oscillator coordinates may be considered to be classical if
\begin{math} \hbar \omega_k \ll k_BT \end{math} in which case
the equipartition theorem asserts that
\begin{equation}
\overline{Q_kQ_{k^\prime }}=
\left(\frac{k_BT}{\omega_k \omega_{k^\prime }}\right)
\delta_{kk^\prime }
\label{mt5}
\end{equation}
wherein the average employs {\em classical statistical mechanics}.
Thus 
\begin{eqnarray}
\overline{\cal V}=\lambda -\Delta {\cal G},
\nonumber \\
\overline{\left({\cal V}-\overline{\cal V}\right)^2}
= \sum_k \omega_k^4 \xi_k^2 \overline{Q_k^2}
= 2k_BT \lambda .
\label{mt6}
\end{eqnarray}
A quasi-classical evaluation of Eq.(\ref{rr7}) is thereby
the Marcus result
\begin{eqnarray}
\tau=\int_{-\infty}^\infty e^{-(i/\hbar)(\lambda -\Delta {\cal G})t }
e^{-(1/\hbar^2)k_BT\lambda t^2} dt,
\nonumber \\
\tau=\hbar \sqrt{\frac{\pi }{\lambda k_BT}}
\exp\left[-\frac{(\lambda -\Delta {\cal G})^2}{4\lambda k_BT}\right],
\nonumber \\
\Gamma = \Omega ^2 \tau \ ;
\label{mt7}
\end{eqnarray}
{\em i.e.} | Eqs.(\ref{intro8}) and (\ref{intro9}) hold true with
\begin{math} \nu =\hbar \Omega^2 \sqrt{\pi/\lambda k_BT} \end{math}.

\begin{figure}
\scalebox {0.5}{\includegraphics{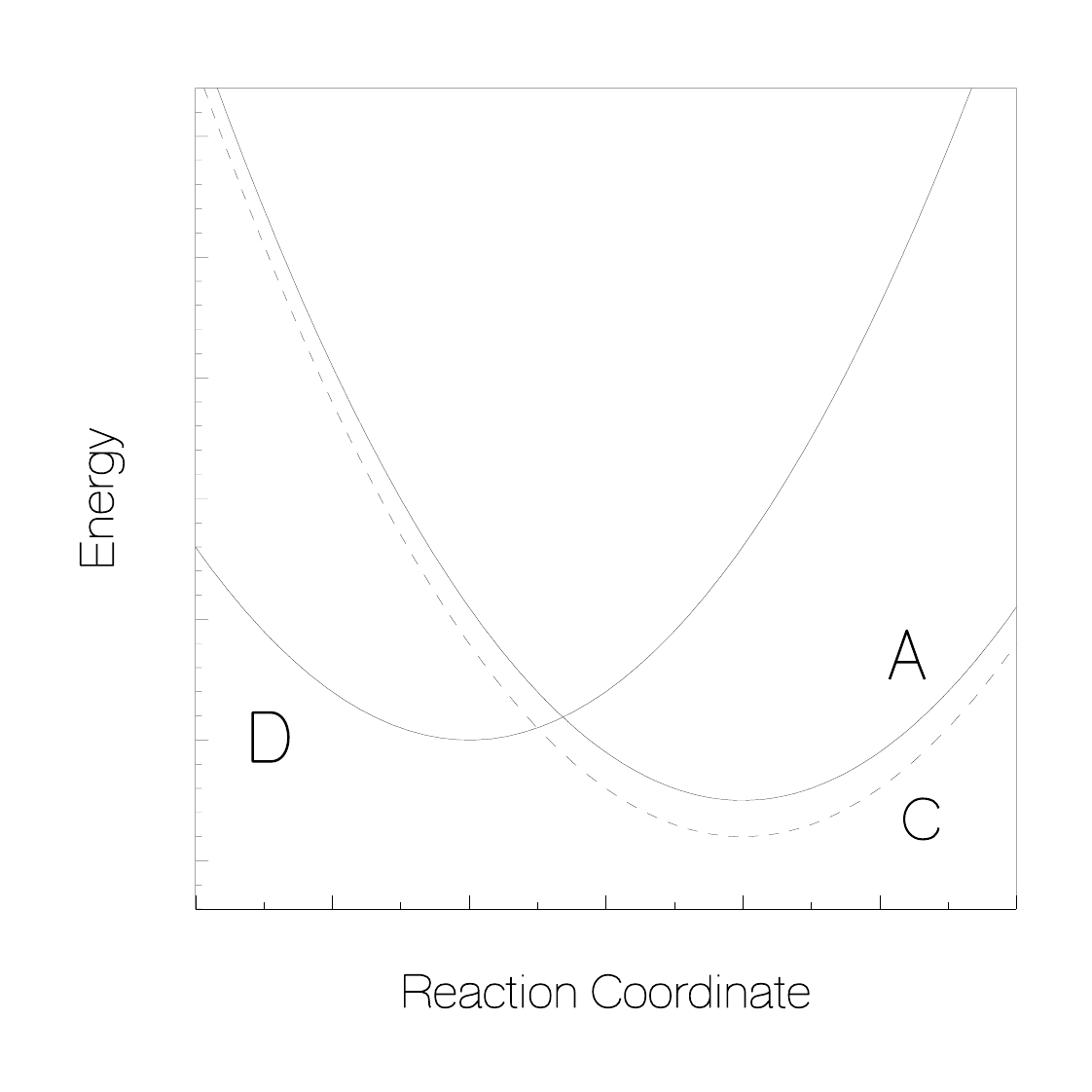}}
\caption{Shown in a schematic fashion are typical diabatic energy
curves for {\it D}-states, {\it A}-states and {\it C}-states describing, 
respectively, the chemical states $DA$, $D^+A^-$ and $(D^+A^-)_{catalytic}$. 
If the catalytic agent increases the chemical binding free energy 
$\Delta {\cal G}$, as in the dotted diabatic curve, 
then the barrier for the reaction is lowered.}
\label{fig2}
\end{figure}

\section{Catalytic Product Binding \label{cpb}}

If the presence of a catalytic agent increases the binding energy of the 
products of a chemical reaction,
\begin{equation}
\Delta {\cal G} \to \Delta {\cal G}+{\cal W},
\label{cpb1}
\end{equation}
wherein 
\begin{equation}
0<{\cal W}< (\lambda-\Delta {\cal G})
\label{cpb2}
\end{equation}
then the Marcus theory barrier free energy \begin{math} {\cal B}  \end{math} for 
the reaction rate 
\begin{equation}
\Gamma =\nu e^{-{\cal B}/k_BT}
\ \ \ {\rm wherein}
\ \ \ {\cal B}=\left[\frac{(\lambda -\Delta {\cal G}-{\cal W})^2}{4\lambda }\right]
\label{cpb3}
\end{equation}
decreases. In fact, if the binding energy \begin{math} {\cal W} \end{math} exceeds 
the upper limit \begin{math}0<{\Delta {\cal G}+{\cal W}}<\lambda  \end{math}, then 
the barrier free energy \begin{math} {\cal B} \end{math} in the reaction rate goes 
to zero. The situation is shown schematically in FIG. \ref{fig2}. The renormalization 
in Eq.(\ref{cpb1}) is equivalent to switching the diabatic curve from 
\begin{math} A \end{math} to \begin{math} C  \end{math} lowering the reaction 
barrier factor \begin{math} {\cal B} \end{math} due to catalytic agents. 
Lowering the barrier free energy increases the reaction rate.

\section{Pure Water Reaction \label{pwr}}

The most simple electron transfer reaction that can take place in pure water 
is, in the notation of Eq.(\ref{intro1}), 
\begin{equation}
H_2O\equiv HOH \ \longrightarrow \ H^+ OH^- .
\label{pwr1}
\end{equation}
How does a metallic surface in contact with pure water increase the reaction 
rate of Eq.(\ref{pwr1}) required to give pure water the well known  
\begin{math} {\rm pH}\approx 7 \end{math} value? It appears that a proton 
is absorbed into the metal leaving an electron outside in an ordered water 
layer above the metallic surface. That extra electron tags onto an 
\begin{math} OH \end{math} bond giving rise to the reaction rate Eq.(\ref{pwr1}) 
in the form   
\begin{eqnarray}
HOH ({\rm liquid}) \ \longrightarrow \ H^+ ({\rm metal})\  OH^- ({\rm liquid})
\nonumber \\
{\rm or\ equivalently} 
\ \ \ \ \ HOH\ \stackrel{\longrightarrow}{_{(metal)}}\ H^+ \  OH^- .
\label{pwr2}
\end{eqnarray}
The reaction in Eq.(\ref{pwr2}) is reversible so that the atoms of the 
catalytic metal play no direct chemical role. Finally, if one inserts small 
spheres of metal into pure water to employ heterogeneous catalysis, then 
the metal spheres will quickly become positively charged by absorbing 
excess protons and ejecting electrons. This is a well-known phenomenon,
perhaps most dramatically displayed in the common undergraduate experiment
in which alkali metals are dropped into water with the ensuing violent reaction
which been the subject of recent detailed experiments\cite{Mason:2015}.

\subsection{The Ordered Water Layer \label{owl}}

Since metal surfaces are in most part hydrophilic, the adjacent water 
layer is ferro-electrically ordered\cite{Siva:2005} as extensively 
studied\cite{Pollack:2008,Pollack:2010,Zheng:2006} by 
Pollack and coworkers\cite{Chai:2009,Chai:2010,Chai:2008} wherein 
the ordered water layer are referred to as an {\em exclusion zone}. What 
is {\em excluded} in the ordered water layer are positive ions and some other 
charged objects. What {\em exists} in the ordered water layer are 
extra electrons as described in the more conventional chemical symbols we 
employ above as negative ions \begin{math} OH^- \end{math}. 
More precisely, the quantitative estimates of the free energy change  
\begin{math} {\cal W} \end{math} requires the physical picture of 
having extra electrons within the ordered water layer. 

The first estimate of the electron chemical potential renormalization 
of electrons within an ordered water domain were made by Preparata 
employing a quantum electrodynamic\cite{{Preparata:1995}} viewpoint 
toward ordered domain formation. The central physical point is as 
follows. If a single water molecule is held together by Coulomb forces alone, 
then then the energy eigenstate wave functions can be chosen as real and 
can be chosen to be eigenstates of the parity operator. The ground state 
wave function for a single water molecule thereby must have a zero mean 
dipole moment 
\begin{math} \left<0\right| {\bf d} \left|0\right>=0 \end{math}. On the other 
hand, the off diagonal matrix elements  
\begin{math} \left<n\right| {\bf d} \left|m\right> \end{math} with 
\begin{math} n\ne m  \end{math} need not vanish. So the physical statement 
that the water molecule has a dipole moment must refer to off diagonal 
matrix elements. If the polarization,  
\begin{equation}
{\bf P}({\bf r},t)=
\left<\sum_{j}{\bf d}_j(t)\  \delta \big({\bf r}-{\bf r}_j(t)\big)\right>, 
\label{owl1}
\end{equation}
is non-zero, then the off diagonal matrix elements of the dipole moment 
\begin{math} {\bf d}_j(t) \end{math}  of the \begin{math} j^{th} \end{math} 
molecule must appreciably contribute to the average in Eq.(\ref{owl1}). Thus, 
in a ferroelectric ordered domain, the internal electronic structure is only 
partly in the ground state and partly in an excited state. This physical 
picture leads to an explanation of why ionic charged particles form in 
electrolytic water solutions, {\em i.e.} why it is so easy to remove an electron 
from a polarized molecule from an excited electronic state. As in all phase 
transitions, the reasons for ferroelectric ordering reside in the collective 
motions of many molecules.

\subsection{Polarizability of a Water Molecule \label{pw}}

Water molecules in the ideal vapor phase have a static 
polarizability\cite{Debye:1928} of the form 
\begin{eqnarray}
\alpha_T = \alpha_\infty +\frac{\mu^2}{3k_BT}\ ,
\nonumber \\ 
\alpha_\infty \approx 1.494\times 10^{-24}\ {\rm cm^3},
\nonumber \\   
\mu \approx 1.855\times 10^{-18}\ {\rm Gauss\ cm^3} .
\label{pw1}
\end{eqnarray}
The value of the dipole moment is often incorrectly associated with a 
mean dipole moment but this must thought through more carefully since  
\begin{math} \left<{\bf d}\right>=0 \end{math}.
The polarizability is in reality 
\begin{equation}
\alpha_T =-\frac{1}{3}\sum_{nm}
\left[\frac{p_n-p_m}{E_n-E_m}\right] \left|{\bf d}_{nm} \right|^2 , 
\label{pw2}
\end{equation}
wherein \begin{math} p_n=e^{(F-E_n)/k_BT} \end{math} is the 
probability of being in state \begin{math} \left|n\right>  \end{math} and 
\begin{math} {\bf d}_{nm}=\left<n\right|{\bf d}\left|m \right> \end{math}. 
From Eqs.(\ref{pw1}) and (\ref{pw2}) it follows that 
\begin{equation}
\frac{\mu^2 }{3k_BT} \approx -\frac{1}{3}\overline{\sum}_{nm} 
\left[\frac{p_n-p_m}{E_n-E_m}\right] \left|{\bf d}_{nm} \right|^2 , 
\label{pw3}
\end{equation}
wherein the restricted sum \begin{math}  \overline{\sum}_{nm}  \end{math} 
requires that energy differences obey 
\begin{math}  \left|E_n-E_m\right|\ll k_BT  \end{math} , {\em i.e.} 
\begin{equation}
\mu^2 = \overline{\sum}_{nm} p_n \left|{\bf d}_{nm}\right|^2 
\equiv \overline{|{\bf d}|^2} < \left<|{\bf d}|^2\right>,
\label{pw4}
\end{equation} 
with the inequality in virtue of the finite but small value 
of \begin{math} \alpha_\infty \end{math}. The root mean square 
fluctuation in the dipole moment 
\begin{math} \mu = \sqrt{\overline{|{\bf d}|^2}}  \end{math} defines 
the so-called {\em experimental} electric dipole moment 
\begin{math} \mu  \end{math} of a single water molecule. 

\subsection{Electron Binding in Ferroelectric Water \label{ebfw}}

\begin{table}
\caption{Work Functions into the Vacuum and into Water}
\label{tab1}
\begin{center}
\begin{tabular}{lrr}
	\hline 
		Metal & \ \ \ \ ${\cal W}_{\rm vacuum} $ &\ \ \ \ ${\cal W}_{\rm water}$ \\ 
	\hline
	\hline	
		Pt & \ \ \ \ 5.5 eV &\ \ \ \ 2.1 eV \\
		Au & \ \ \ \ 5.2 eV &\ \ \ \ 2.3 eV \\
		Cu & \ \ \ \ 4.7 eV &\ \ \ \ 2.1 eV \\ 
	\hline 
\end{tabular}
\par\medskip\footnotesize
Order of magnitude experimental determination\cite{Musumeci:2012}  
estimates of the work required to move an electron from the metal 
into the vacuum from the metal into and a polarized layer of water. 
\end{center}
\end{table}

To remove an electron from an isolated water molecule that is initially 
in the ground state requires an energy of 
\begin{math} \varphi_0 \approx 12.6\ {\rm eV} \end{math}. 
To excite an electron from an isolated water molecule that is initially 
in the ground state and finally in the first excited state requires an 
energy given by \begin{math} \varphi_1 \approx 10.2 \ {\rm eV} \end{math}. 
The difference \begin{math} \phi = \varphi_0 - \varphi_1 \end{math} is 
thereby 
\begin{equation}
\phi \approx 2.4\ {\rm eV}.
\label{ebfw1}
\end{equation}
As a first approximation, the work function to bring an electron from a metal 
into a ferroelectric ordered layer of water is 
\begin{math} {\cal W}_{\rm water}\approx \phi \end{math}. This work function is 
in qualitative agreement with data listed in TABLE \ref{tab1}. The rates of the 
pure water reaction Eq.(\ref{pwr1}) in the neighborhood of the metallic surface 
is thereby such that the barrier would be virtually eliminated in virtue of 
\begin{math} \lambda \ll {\cal W}_{water} \approx \phi \end{math}.

We note that metal catalyzed reactions in water are of significant current
interest, especially with aim towards ``greener'' syntheses \cite{green-chem}.

\section{Metal-Insulator Catalysis\label{mic}}

\begin{figure}
\scalebox {0.5}{\includegraphics{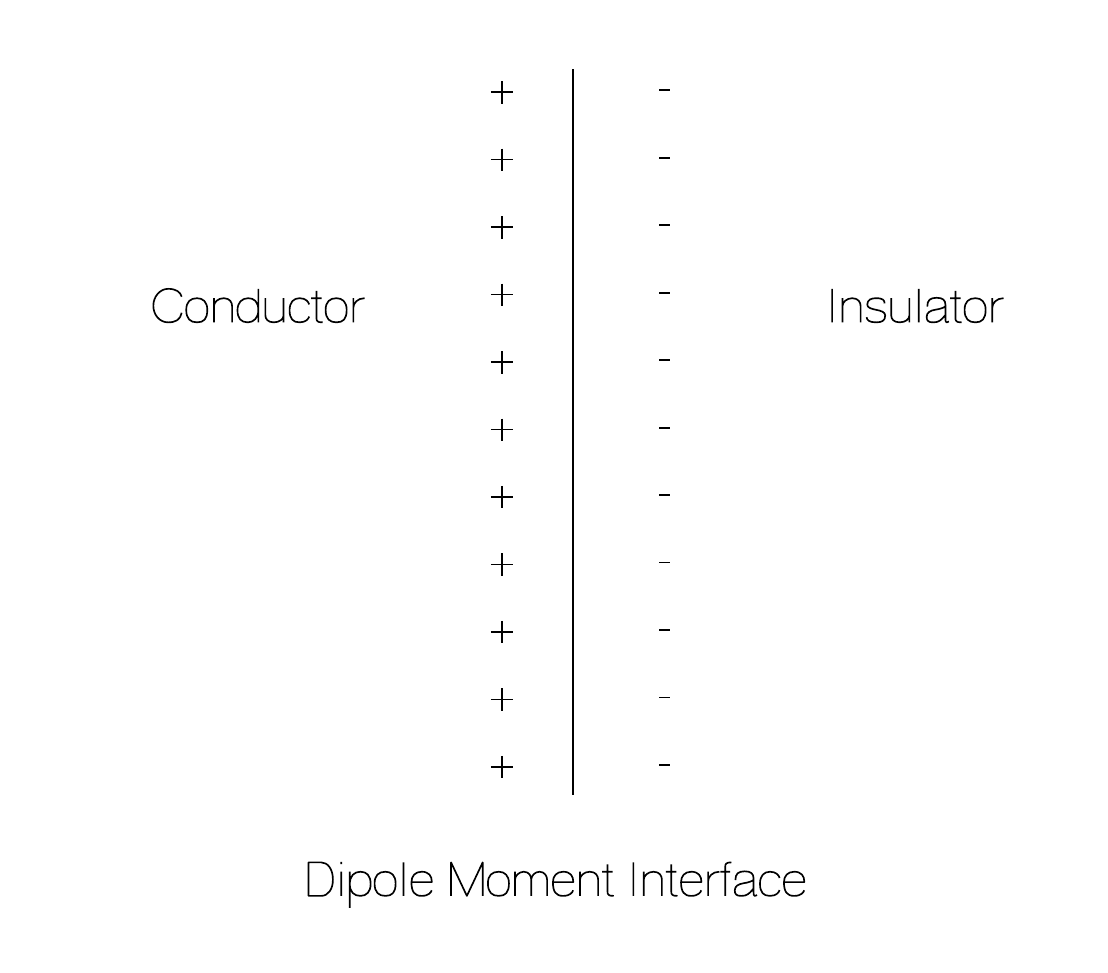}}
\caption{Shown is a schematic view of the metal-insulator interface. The vacuum 
in this regard is a possible example of an insulator. The electrons can lower their 
kinetic energy by leaking out of the metal and into the insulator creating a dipole 
moment per unit area as pictured above. The interface dipole moment points from the 
insulator into the metal.}
\label{fig3}
\end{figure}

For the case of a metal catalytic agent in contact with an insulator, 
({\em e.g.} hydrogenation of oils in the presence of a nickel catalyst, the Haber-Bosch
synthesis of ammonia from hydrogen and nitrogen catalyzed by iron, or any of countless
organic reactions involving nonpolar solvents catalyzed by metals\cite{organic-text}), one may apply the
Lippmann-Gibbs equation\cite{Landau:1984} to  the interface between the two phases.
Our sign convention is that the electronic charge is negative, 
\begin{equation}
e=-|e|,
\label{mic1}
\end{equation} 
so that the constant electronic chemical potential 
\begin{math} \tilde{\mu } \end{math} obeys 
\begin{equation}
\tilde{\mu }=\mu -e\Phi =\mu +|e|\tilde{\Phi } 
\ \ \ \Rightarrow \ \ \ \mu-\tilde{\mu }=e\Phi =-|e|\Phi  
\label{mic2}
\end{equation}
wherein \begin{math} \mu  \end{math} is the local chemical potential and 
\begin{math} \tilde{\Phi } \end{math} is the electrostatic potential, {\em i.e.} the electrostatic 
field is given by  
\begin{equation}
{\bf E}=-{\bf grad}\tilde{\Phi }. 
\label{mic3}
\end{equation}
Consider the metal-insulator interface shown in FIG. \ref{fig3}. The the x-axis 
pointing positively normal to the interface from the metal into the interface, 
one may invoke the Poisson equation, 
\begin{equation}
-\frac{d^2 \tilde{\Phi }(x)}{dx^2}=4\pi \rho (x),   
\label{mic4}
\end{equation}
to compute the dipole moment per unit area \begin{math} \tau \end{math} 
employing 
\begin{eqnarray}
\tau =\int_{-\infty}^\infty x \rho(x)dx, 
\nonumber \\ 
4\pi \tau = -\int_{-\infty}^\infty x
\ \frac{d^2 \tilde{\Phi }(x)}{dx^2}\ dx, 
\nonumber \\ 
4\pi \tau = \int_{-\infty}^\infty 
\ \frac{d \tilde{\Phi }(x)}{dx}\ dx=
\tilde{\Phi }(\infty )-\tilde{\Phi }(-\infty ), 
\nonumber \\ 
4\pi \tau = -\Phi \equiv V_{\rm contact} < 0,  
\label{mic5}
\end{eqnarray}
wherein \begin{math} V_{\rm contact}  \end{math} is the contact voltage 
across the interface when viewed as a possibly non-linear capacitor. The 
length scale ``\begin{math} \infty \end{math}'' refers to a length much 
larger than the interface dipole moment length \begin{math} L_T \end{math} 
described by Eqs.(\ref{lge1}) and (\ref{lge2}) below. 

\subsection{Lippmann-Gibbs Equation \label{lge}}

If \begin{math} \sigma  \end{math} denotes the free energy per unit area 
(surface tension) of the interface, \begin{math} s \end{math} denotes 
the entropy per unit area and \begin{math} \varpi  \end{math} denotes 
the charge per unit area on the charged sheets shown in FIG. \ref{fig3} when 
viewed as a non-linear capacitor, then the thermodynamics of such a capacitor 
is described by the Lippmann-Gibbs equation 
\begin{equation}
d\sigma =-s dT-\varpi d\Phi .
\label{lge1}
\end{equation} 
The capacitance per unit area is thereby 
\begin{equation}
K_T=\left(\frac{\partial \varpi }{\partial \Phi }\right)_T 
\ \ \ \ \Rightarrow
\ \ \ \ 4\pi K_T=\frac{1}{L_T}\ ,
\label{lge2}
\end{equation} 
wherein \begin{math} L_T  \end{math} describes the effective distance between 
the charged sheets pictured in FIG. \ref{fig3}. The dipole moment per unit area 
\begin{math}  \tau  \end{math} is in a very crude order of magnitude given by 
\begin{math} \sim  \varpi L_T  \end{math}. 

\subsection{Marcus Electron Energy Renormalization \label{meer}}

The work function to remove an electron from an insulating layer coating the 
metal may be written as 
\begin{eqnarray}
{\cal W}_{\rm insulator}=\mu =\tilde{\mu }+e\Phi =\tilde{\mu }-|e|\Phi ,
\nonumber \\ 
{\cal W}_{\rm insulator}={\cal W}_{conductor}-|eV_{\rm contact}|, 
\nonumber \\ 
{\cal W}_{\rm insulator} < {\cal W}_{\rm conductor}\ . 
\label{meer1}
\end{eqnarray}
The coating of a metal with an insulator lowers the work function. As a result, 
the free energy \begin{math} \Delta {\cal G} \end{math} of an electron transfer 
reaction is raised via 
\begin{equation}
\Delta {\cal G}\to \Delta {\cal G}+|eV_{\rm contact}|.
\label{meer2}
\end{equation} 
Since the contract potential \begin{math} |V_{\rm contact}| \end{math} is of the 
order of volts and the reorganization energy \begin{math} |\lambda /e| \end{math} 
is of the order of less than than a tenth of a volt, under the 
renormalization Eq.(\ref{meer2}) the barrier energy is eliminated. This allows for the 
explanation of the catalytic properties of the metal. Specific differences between different
metal catalysts requires taking into account factors such as absorption, desorption, and diffusion issues
at the surfaces, details of the electronic structure of the metals, and surface preparation
and is beyond the scope of this paper which is to simply explain why metal catalysis
can be so spectacularly successful. We hope to return to some of these other matters
in a future publication.

\section{Conclusion \label{conc}}

The Marcus theory of electron transfer organic reaction rates has been applied 
to the catalytic properties of contact with bulk metals. The electric fields 
normal to the interface and the resulting contact potentials reduce or even eliminate the barriers 
and thereby strongly increase the kinetic rates of reactions. The energy inequality 
required for the catalytic effect, requires that the contact potential is large on 
the scale of the Marcus renormalization energies. 

\section*{Acknowledgments}

J. S. would like to thank the United States National Science Foundation for support under PHY-1205845.

\end{document}